\documentclass[sigconf]{acmart}

\AtBeginDocument{%
  \providecommand\BibTeX{{%
    \normalfont B\kern-0.5em{\scshape i\kern-0.25em b}\kern-0.8em\TeX}}}

\copyrightyear{2021}
\acmYear{2021}
\setcopyright{acmlicensed}\acmConference[AIES '21]{Proceedings of the 2021 AAAI/ACM Conference on AI, Ethics, and Society}{May 19--21, 2021}{Virtual Event, USA}
\acmBooktitle{Proceedings of the 2021 AAAI/ACM Conference on AI, Ethics, and Society (AIES '21), May 19--21, 2021, Virtual Event, USA}
\acmPrice{15.00}
\acmDOI{10.1145/3461702.3462604}
\acmISBN{978-1-4503-8473-5/21/05}

\begin{document}
%\fancyhead{}

\title[We Haven’t Gone Paperless Yet]{We Haven’t Gone Paperless Yet: Why the Printing Press Can Help Us Understand Data and AI}

\author{Julian Posada}
\email{julian.posada@mail.utoronto.ca}
\orcid{0000-0002-3285-6503}
\affiliation{%
  \institution{University of Toronto}
  \country{Canada}
}

\author{Nicholas Weller}
\email{nweller@ucr.edu}
\orcid{0000-0002-2198-5625}
\affiliation{%
  \institution{University of California, Riverside}
  \country{United States}
}

\author{Wendy H. Wong}
\email{wendyh.wong@utoronto.ca}
\affiliation{%
  \institution{University of Toronto}
  \country{Canada}
}

\begin{abstract}
  How should we understand the social and political effects of the datafication of human life?  This paper argues that the effects of data should be understood as a constitutive shift in social and political relations.  We explore how datafication, or quantification of human and non-human factors into binary code, affects the identity of individuals and groups.  This fundamental shift goes beyond economic and ethical concerns, which has been the focus of other efforts to explore the effects of datafication and AI.  We highlight that technologies such as datafication and AI (and previously, the printing press) both disrupted extant power arrangements, leading to decentralization, and triggered a recentralization of power by new actors better adapted to leveraging the new technology.  We use the analogy of the printing press to provide a framework for understanding constitutive change.  The printing press example gives us more clarity on 1) what can happen when the medium of communication drastically alters how information is communicated and stored; 2) the shift in power from state to private actors; and 3) the tension of simultaneously connecting individuals while driving them towards narrower communities through algorithmic analyses of data.
\end{abstract}




\maketitle

\section{Introduction}
We are currently in the era of datafication.  Datafication, as we use it here, is the process by which the world is processed, quantified, and stored digitally \cite{Mayer-Schonberger2014, couldry_deconstructing_2018, Mejias2019} and converted into binary code. One important development of datafication is the proliferation of AI technologies. Without the rise of “Big Data,” data-intensive machine learning techniques would not have been able to make the enormous strides that they have and continue to do. With this increase in data production, analysis, and access, however, we have at the same time struggled to develop strategies to govern data, regulate AI, and articulate the social and political effects of datafication and artificial intelligence. Already, a mounting pile of quandaries have arisen about the scope and ethics around data with very few good answers.  In this paper, we show that comparing datafication to the dynamics of the printing press’ development in Europe can help to anticipate what might come from datafication and AI.

We argue in this paper that fundamental shifts in information technology – such as our current age of datafication – are characterized by the tension of a sequential decentralization of power enabled by the technology and recentralization of power as actors adapt to the new possibilities. Datafication has given private economic actors the ability to surveil individuals in ways we used to typically understand as belonging to states only \cite[19]{fourcade_seeing_2017}. These changes in power are accompanied by changes in the identity of individuals and groups, what we label \textit{constitutive} shifts. As with the printing press in Europe, datafication opened up the floodgates of what we know, and how quickly we can know it.  Literacy as a minimal qualification was replaced with access – to devices, to platforms, to the Internet.  With this came an emancipatory sense, as in the case of the printing press’ break of the Church’s stranglehold on words and reproducible knowledge. 

Datafication, however, has also ushered in a new centralization of power, as those who developed the means to cull data most efficiently and widely have come to disproportionate amounts of power. Social media firms, such as Facebook and Twitter, are designed for gathering data \cite{deibert_reset_2020}, and they have come to dominate how we think about freedom of expression. This came to the fore with their decisions to allow Donald J. Trump and his allies to continue spreading misinformation throughout his presidency, and their choices to finally cut him off after the January 6, 2021 Capitol riots.  Search firms, such as Google, and e-commerce companies, such as Amazon, determine how people see the world and to what they have access. These dynamics have intensified since the COVID-19 pandemic began.

To understand the extent to which datafication and AI might reshape life as we know it, we can look to analogies with shifts wrought by the printing press. The printing press fundamentally changed life in early modern society and politics in three constitutive ways that are relevant to the current day.  First, it moved society from one of oral orientation to one of visual orientation.  It made communication individually contemplative, rather than shared. It also committed ideas to paper as a medium \cite{deibert_parchment_1997}, which has its own organizational logics regarding how information is conveyed. Second, it shifted power from the Church and state as the keeper of knowledge to private actors.  It realigned economic interests and power as well, empowering publishers and printers \cite{Yale2015, Ruud1980}. Third, it allowed for greater possibilities for human imagination as “communities” with the growth of newspapers and the novel. The archive, as a repository for select documents and publications, created for some social purpose, then became a source of information, truth, and power \cite{Yale2015}.

The printing press is the most relevant technological change to look for as an analogy because it is changed the production of information, its reception and the way it is used, and its storage. Like the printing press, datafication fundamentally alters our social and political relationships because they store and share data in a much more more efficient format. This allows for the massive gathering of data.  AI is the means by which we process and organize those data;  AI also trains on that archive of data \cite{Jo2020}.  First, datafication and AI have moved us to a society of archiving and prediction, of the action taking place on servers and through algorithms indiscernible to the users. It is both more public on its face, and yet private in the way that the data are stored and used. Second, these forces have shifted power from states to private actors that develop and own the algorithms, and it is these actors who invent the categories of data that are to be collected, and implement the algorithms to collect and analyze data from users. While it is true that individuals have more access to data and information, it is also true that data are being collected from individuals in a range of ways that are both voluntary and involuntary \cite{veliz_privacy_2020}. Third, it has both created more opportunities for communities to grow while also providing the opportunity for deeper fractures as algorithms shuffle people into different “filter bubbles.” Algorithms have become arbiters of truth and power by determining what data are viewed, by whom.

In Section 2, we review two prominent, alternative ways to think about the effects of datafication and AI, through their economic and ethical implications.  In Section 3, we explore the three constitutive ways in which the printing press’ effect on social and political structures in Europe hold key lessons for us as we face a widespread shift in how information is conveyed and communications are conducted. In this section, we particularly focus on the simultaneous decentralization forces of the technology that is coupled with the countervailing tendency for actors to centralize access and influence. Section 4 concludes with some proposals on how to integrate insights from the constitutive changes wrought by technological change to policymaking.

\section{Ways of Looking at Datafication and AI}

Datafication has many different definitional enhancements, depending on the author, but its main point often focuses on what Mejias and Couldry call “[a] contemporary phenomenon which refers to the quantification of human life though digital information, very often for economic value” \cite[1]{Mejias2019}. The importance of datafication, however, is not just in the generation of initial data, but the value that can be extracted from that data \cite{Mejias2019}. Put differently, the value of data lies in both collecting and drawing inferences from that data. Thus, datafication would not have nearly as much influence without the capabilities of AI to leverage the data into predictions.

\subsection{Economics}

The economics of datafication have focused on the financial imperatives and gains for those who create data from observable reality \cite{sadowski_when_2019,van_dijck_datafication_2014}. Terms such as “surveillance capitalism” \cite{zuboff_age_2019} permeate the public imagination to convey how companies are harnessing the data we generate to create wealth for themselves while also controlling our access to information. Others, such as Fourcade and Healy emphasize the stratification and classification of consumers, through the analysis of the data they generate, and the economy of moral judgment that follows from categorizing “good” and “bad” consumers \cite{fourcade_seeing_2017}.  

The collection and flows of these datasets have been described by Jo and Gebru as influenced by a \textit{laissez-faire} attitude from practitioners and tech companies with little care about the social implications of the data they collect \cite{Jo2020}. This culture is related to the influence of the idea that data is a nonrival good (not depleted by consumption) and a by-product of economic activity \cite{Jones2020, Varian2019}. Some claim that data can be “owned” privately by firms or consumers and transacted in markets \cite{Ichihashi2019}. Put into practice, this conception of data as a market good creates a pervasive environment for the collection and processing of data to train machine learning algorithms that do not account for the rigorous and ethical processes of other data-intensive sectors and fields such as healthcare and social science disciplines \cite{Paullada2020}.

Another influence of economic and business thinking that permeates the datafication process required for AI is the desire to create scalable systems. Scalability is the expansion of a system without the modification of core features \cite{Tsing2012}. This notion of economic growth through the exploitation of nonrivalrous datafied goods has become fundamental in the startup culture that fuels some recent innovations in artificial intelligence. For Hannah and Park, this scalable thinking permeates the development of predictive systems with the following set of assumptions: that these systems are ethically desirable, depend exclusively on the quantification of human experience, and that a system of any scale requires only a limited number of core functions \cite{Hanna2020a}. In practice, scalable thinking would mean that a system developed in the Silicon Valley could potentially grow to be implemented in other parts of the world without changing its core features, being only necessary to adapt superficial characteristics to the necessities of new markets. 

The economic perspective on datafication does not take into account how these data-driven avenues pursued for market advantage fundamentally alter the way that humans interact. The view from here emphasizes why firms choose certain types of data collection and analysis with a simplified (if any) account of the social and political effects of these actions.

\subsection{Ethics}

There is also the political and ethical perspective on datafication, which can be captured best by the phenomenon of “dataveillance.”  Dataveillance can be defined as “the name for the disciplinary and control practice of monitoring, aggregating, and sorting data” \cite[124]{raley_dataveillance_2013}.  Unlike surveillance, which monitors the physical being, “dataveillance watches the shadow that the person casts as they conduct transactions, variously of an economic, social or political nature” \cite{clarke_dataveillance_2017,clarke_information_1988,clarke_information_1988}. Thus, technology is allowing for greater levels of intrusions upon individual lives, at times justified by security needs.

Since the applications of AI have become too important to be ignored by governments, institutions, and corporations, there has been an increasing interest in how the deployment of this technology would benefit society. In this context, discussions about “ethical AI” have permeated a sector of academia and policy for the past years. Carly Kind, director of the Ada Lovelace Institute, has recently identified three different waves in these discussions, each of them influenced by major academic disciplines \cite{Kind2020}. Calling them “waves” does not mean that research from these frameworks has stopped. Instead, new disciplines and perspectives have complemented these different research streams.

The first wave, heavily influenced by philosophy, has focused on general ethical principles that AI systems should follow. These principles came primarily from dominant industry and governmental actors and converged in ideas such as transparency and fairness \cite{Jobin2019}. However, these principles remained abstract and with no clear consensus on how to implement them, leaving meaningful discussions on policy and regulation behind \cite{Calo2017}. Furthermore,  these discussions have been criticized as “ethics washing” because of their lack of applicability \cite{Hao2020a} and funding from “big tech” \cite{Abdalla2020}.

A second wave tries to resolve some of the dilemmas from the first technically.  Computer science has focused on how algorithms can become “fair” and “unbiased,” exploring ways to collect more diverse data while solving algorithmic discrimination in mathematical models. Research on fairness from a computational perspective has focused on developing models that address the difference in “performance” between categories such as “sub-populations” and “individuals” \cite{Chouldechova2018,Corbett-Davies2018, Dwork2012,Torralba2011}. However, while it is essential for these mathematical models to address the data differently for categories or sub-groups, thinking about “fairness” purely in statistical and mathematical terms does not fully address AI’s societal implications. These predictive systems are sociotechnical in nature \cite{Kind2020}.  An understanding of social contexts and relations that go beyond computational models is necessary for constructing and implementing algorithms \cite{Hadfield-Menell2019a, Denton2020,Miceli2020}.

The third wave is much more heavily influenced by the social sciences. It explores the power relations behind the development and deployment of artificial intelligence \cite{Kalluri2020,Mohamed2020,Birhane2020}. A subset of this research on AI from a sociotechnical perspective looks at the data supply chains that allow the creation of machine learning systems from the datafication process for input, training, and feedback data \cite{Newlands2021,Agrawal2018}. Beyond the widely discussed issues of privacy in data collection for consumers from the social sciences \cite{Madden2017,Arora2019} and computer science \cite{Hildebrandt2019}, these data supply chains also involve a different array of actors, including data workers \cite{Posada2020a}, data brokers \cite{Anthes2015}, and other institutional and governmental actors.

This final wave draws attention to some of the directly-observable trends that datafication and AI have brought on. We start from this perspective, and theorize more broadly about the effects of these technologies. Using the printing press’ history as a guide, we “widen the lens” to consider the findings from the third wave for bigger, constitutive effects of datafication and AI.

\section{Why Datafication and AI are Analogous to the Printing Press and Archive}

\subsection{An Archiving and Predicting Society}

Datafication provides the raw material for AI to do what it does best: provide predictions \cite{Agrawal2018}. The data for these predictions, however, are coming from participants in a relationship of unequals between the companies that collect the data and make predictions, and the individuals whose data are being gathered and analyzed. As consumers, we have participated in helping the rise of Big Data and the greater accuracy of machines \cite{nourbakhsh_ai_2020}. Giving away this information has the effect of transferring power to those who archive the data, but it also has ramifications on individual agency. For example, Cathy O’Neil provides an easily-accessible laundry list of ways that discrimination and bias are replicated in the algorithms that determine what jobs we are qualified for and whether we get home loans \cite{oneil_weapons_2017}. Ruha Benjamin has powerfully demonstrated how these dynamics reinforce existing racial inequities in what she calls “the new Jim Code” \cite{benjamin_race_2019}. However, much of our participation has not occurred in a situation of transparency, awareness, or even the ability to easily disentangle ourselves even if we were aware of the situation.

As with the printing press, where the medium upon which ideas were communicated also generated expectations of what knowledge “looked like,” datafication has generated expectations about the medium in which communications take place.  These communications media are data-intensive.  In 2019, 4,497,420 Google searches were performed and 55,140 photos were posted on Instagram every minute \cite{Domo2020}. This is based on just over half of the world’s human population being online. All of these data are providing new ways in which information is communicated and absorbed by users of the Internet. And these data generate more data about what users want and do not want, how ideas trend, and what kinds of concerns are of the moment.  These are all reflections of individual taste, personality, preference – in short, the data give the collectors of that data a good sense of “who” their users are \cite{veliz_privacy_2020}.  They can also give away what kinds of relationships people have to one another, and the quality of those relationships, something that has been called “relational big data” \cite{levy_relational_2013}.

The strength of the prediction is only as good as the data the algorithm trains on.  Our society is increasingly geared towards the collection of data, for the purposes of improving prediction through AI.  These are quite frequently tied to monetary imperatives, because the entities doing the collecting are mostly companies. Governments have also done their fair share of using data, but often that data is provided by companies with technologies to gather data \cite{Marczak2015}.

The orientation towards sharing in public fora requires that users interact with data-gathering technology. It also has the effect of hiding where all the data go after we have provided them, whether this is our behavior online as tracked by cookies, or the photos we upload and share on social media. One particular example is with facial data, and the use of facial recognition technology.  Though not a new technology – it started in the 1960s \cite{Raviv2020} – it has become commonplace, showing up in apps like Facebook or iPhoto, in smart home products like Google Nest, and in car safety applications like the Subaru DriverFocus Distraction Mitigation System.  It is used in airports, on city streets, at Taylor Swift concerts \cite{Canon2019}, in retail stores, and more \cite{Brown2019}. It is also used by police to do their jobs. But police had not had access to the numbers of faces and the volume of facial data until Clearview AI stepped onto the scene. Clearview claims it has amassed 3 billion face images, all scraped from websites such as Facebook, YouTube, and Venmo \cite{Hill2020b}. Clearview works with over 2400 law enforcement agencies \cite{Lopatto2020}, and helps companies with their security too \cite{Hill2020b}. These data – the archives of faces – are sometimes voluntarily gotten in the first instance, as in the case of Clearview’s database, which was taken from photos that users had voluntarily uploaded, albeit for some other purposes. Once in the face databases, it is not clear how one might get out, if ever, creating questions about to whom a face’s data belong \cite{wong_as_nodate}.

An important issue with facial recognition technology is how individuals within a datafied society consent to the collection, archiving, and use of the data they provide.  There is not a good way to think about this from a governance perspective, either in terms of how one might consent to the myriad items that app and service creators embed in their terms and conditions; or how devices such as FitBit are collecting vital information that at worst constitute surveillance and social conditioning, but at best are just invasive \cite[21-26]{frischmann_re-engineering_2018}; or from the view of how someone might be able to just forfeit their right to privacy because of the way datafication renders such a right quite difficult to protect \cite{sinha_real-property_2018}. 
Furthermore, even if individuals did consent, the biases baked into the way our data are collected and the algorithms are written continue to vex the effectiveness and accuracy of technologies such as facial recognition  \cite{benjamin_race_2019,Chokshi2019,Buolamwini2018,noble_algorithms_2018}. Recent reports of false arrests of Black men, based on facial recognition technology, for example, highlight how the technique replicates existing discrimination and subjugation of racialized groups \cite{Hill2020c}. In general, studies found that facial recognition technology was just not very accurate in matching faces in real time to those in the database.  In one set of trials in the UK, the accuracy rate was reported to be 19.05\% \cite{Burgess2019}.

Not surprisingly, our institutions have not caught up with the sheer volume of datafication, or the technical advances made to improve AI technologies to assess that data.  We are being asked to reevaluate how we live our lives, understand our rights, and know where our data are going, when the entire system of datafication of communications, social interactions, and knowledge transfer are happening on a completely new medium that is not the same as the ones with which we have more intimate knowledge, such as paper. The move to datafication and the use of AI to move through those vast stores of data can be better understood if we take the view from the transition from hand-scribing to printing press.  Where prior to the printing press, the word was literally sacred because very few could read \cite[49-52]{deibert_parchment_1997}, the same applies to data today.  Data have been mysterious and the domain of technical specialists, because they have been the ones creating the tools that are widely used today.  The bait-and-switch has been that individuals have all been allowing their lives to be datafied, and very much giving away some part of themselves in this technological transition.  Thus, it seems quite reasonable to say that everyone alive today has a very real stake in how human lives are being converted into datafied forms.

\subsection{Changes in Power Structures Wrought by Technology}

The development of the internet and datafication has been compared to the printing press by a multitude of scholars. New communications media not only establish the forms in which individuals communicate, but also create new ways to interact and the possibilities for new kinds of social ties \cite{thompson_media_1995}. Unlike other media innovations, such as the telegraph or the related invention of the telephone, the printing press produces communication on a mass level, intended for a wide audience, and serves medium of communication that fundamentally shifted the way that human being relate to one another in time and space. Datafication, as described above and discussed below, has very much these same qualities. But unlike the printing press, or telephone and telegraph, datafication’s link to AI has the additional quality of culling data directly from all who use the technology. 

Research on the social and political effects of the printing press are not new. Eisenstein argues that the effects of the printing press were not fully appreciated, because scholars had focused mostly on how the press affected the dissemination of ideas \cite{eisenstein_printing_1979}. She demonstrates that the press led to fundamental changes in what people thought and how they thought about the world in such profound ways that it influenced the Reformation, the Renaissance, and the Scientific Revolution. 

The printing press has also been connected to other massive changes in social and political organization and the distribution of power. In \textit{Imagined Communities}, Benedict Anderson argues that the newspaper led to the rise of nationalism, as it made it possible for disparate individuals to share a common language, time, calendars, and other factors that built a shared community \cite{Anderson1983}. Deibert \cite{deibert_parchment_1997} also focuses on the role that the printing press had on power relations to argue that it was an important cause in the development of the modern nation-state system, and he argues that new technologies are likely to disrupt current power relations. 

More specific changes in power distribution were wrought by the printing press.  In the Russian Revolution context, the printing press created a group of workers who were important to the structure and organization of society but who did not exist prior to the creation of the printing press \cite{Ruud1980}. In fact, printers were among the first to organize.  Their role in the production of reading materials gave them a unique skillset both for the access to printing and also in their high levels of literacy.  They were instrumental in undermining censorship laws early in the 20th century.

From Ruud’s history, we can see that the printing press created a new class of employees that went from non-existent to politically powerful in a short period of time. In the early 1900s Russian printers took the lead in the development of a free press and therefore this occupational group which had not even existed 50 years earlier was now leading radical changes in Russian society. Ruud argues that “the modern printing press was itself a powerful – but so far little recognized – ‘agent’ of political change during the Revolution of 1905 in Russia” \cite[395]{Ruud1980}. The printing press was not developed to aid in revolutionary political change, but profoundly changed how power was distributed.  The case demonstrates how technologies can have unanticipated consequences in terms of usership, skill development, class creation, and group mobilization, all of which can contribute to shifts in power.  In the case of Russia, this new class of printers were able to halt the repressive policies of the czarship and foment support for the growing revolution.

The Russian example demonstrates how a new technology created a class of private actors that changed society. Although the process is different, datafication has likewise empowered individuals and businesses that have the potential to change society and politics. Corporate actors have become key players in the amassing of data generated by individuals interacting with technology.  The biggest players in AI are just nine firms based in China and the US \cite{webb_big_2019}.  The voracious data appetites of social media \cite{deibert_reset_2020}, search, and platform providers have been well-documented \cite{zuboff_age_2019}.  Their imperative to keep their user base (and keep growing) is understood \cite{karppi_disconnect_2018}, as without users, these firms would not have their advantageous positions in data access.  

As with the printing press, the advent of datafication as a technology has ushered in actors who have been able to capitalize on the environmental shift that privileges data collection, archiving, and prediction.  The current US Department of Justice anti-trust suit against Google speaks to the ways that the company controls search renders it the ubiquitous keeper of information.  The outcome of Google’s data collection, archiving, and predicting efforts is that they effectively control the public record \cite{noble_algorithms_2018}.  However, even as it serves a public purpose in monopoly fashion, Google responds to market incentives: its advertisers \cite[36]{noble_algorithms_2018}. Together with social media giants such as Facebook, Tiktok, and Twitter and app gatekeepers such as Apple and Amazon, they effectively gatekeep what we know, how we know it, and if we know it.  This is analogous to how control over the printing press limited what information was received, in what format, and by whom. For example, European colonizers in Latin America regulated printed materials in order to control the spread of information and ideas. Later on, clandestine presses were fundamental in spreading revolutionary ideas that sparked the wars of emancipation in the early 19th century \cite{RoldanVera2013}.

This stands in contradiction with much early discussion of the internet, which talked about it as a democratizing force in its ability to diffuse power away from the elites and to mass citizens in a way that appears analogous to the printing press from centuries earlier \cite{Dewar1998}. While clearly information is more accessible than ever before for ever larger segments of society, there is a countervailing trend associated with the collection and analysis of massive amounts of data.  “Big Data” has created new economic actors: the (mostly) firms that amass and analyze these data. Just as the press moved power and information away from government and religious institutions and into the hands of private actors such as printers, datafication and AI have given certain kinds of technology companies a leg up (so-called “Big Tech”).  Perhaps most importantly, the more we use these technologies, the more powerful they get \cite[14]{Pasquale2015b}.  The users use Google search or scroll through their Twitter feeds, the more data these companies collect, the more analysis they provide, and they more they understand about individuals and their users as a collective.

Unlike today’s internet landscape, characterized by the prevalence of private enterprises and dominated by a handful of them, sometimes called “Big Tech” or GAFAM (Google, Apple, Facebook, Amazon, and Microsoft), the early infrastructure of the Advanced Research Projects Agency Network (ARPANET), the ancestor of the internet, and subsequent technical developments before its privatization, were financed, developed, and maintained by public entities and resources. The early infrastructure was a public good, financed by the United States government through its military and developed primarily by universities and not-for-profit research centres across the country \cite{Smyrnaios2017}. Similarly, other developments that became fundamental for today’s internet, such as the TCP/IP protocol and the World Wide Web, were also conceived by researchers in international and national publicly-funded institutions such as the European Organization for Nuclear Research (CERN) \cite{gillies2000web}.

The commodification and privatization of the Internet occurred progressively in the 1980s and early 1990s followed the neoliberalization under Reagan in the United States \cite{gillies2000web,harvey2007brief}. In the case of the Internet, this shift challenged the idea that single networks supervised by a national regulatory body were required to protect the public interest \cite{Smyrnaios2017}. The following years after the privatization of the internet saw the emergence of today’s internet market and the birth of contemporary Big Tech companies, despite the dotcom bubble of the early 2000s. Today’s significant players profited from early setbacks, from the purchase of patents from extinct or failing businesses and to their acquisition and integration into major corporations \cite{foroohar2019don}.

Several economic, political, and social conditions coalesced that brought about Big Tech. These conditions were produced by federal regulators not addressing the increasing size and power of these companies, their steady accumulation of capital, and their tax avoidance practices \cite{Smyrnaios2017}. Along with these characteristics, as mentioned above, major technology companies also based their growth on their reliance on non-rivalrous goods \cite{Jones2020}, the reduction of production costs, notably through the exploitation of labour \cite{Casilli2019a}, and a “winner-take-all” form of competition characterized by the acquisition of competitors or their suppression through coercion and other means \cite{Smyrnaios2017,foroohar2019don}. These dynamics make oligopolies more likely. Globally, the strategy of Big Tech has also been characterized by its dependence on scalability, where their growth is not dependent on major changes to their operations and business models \cite{Hanna2020a}.

Another critical component of the concentration of wealth and power in contemporary datafication is Big Tech’s reliance on the socioeconomic and infrastructural model of the platform to coordinate markets, social relationships, and information \cite{Casilli2019}. In this context, platforms are “(re-)programmable digital infrastructures that facilitate and shape personalised interactions among end-users and complementors” \cite[3]{Poell2019}. Thus, as platforms, Big Tech controls the informational and material exchanges between the many actors in its markets, for instance, by regulating information flows through algorithms with end-users \cite{Burrell2016,kitchin_thinking_2017} and the interplay between platforms and back-end users \cite{Nieborg2018e}. Big Tech sits at the nexus of datafication, having created the systems that others must use, and the monitoring and maintaining of those systems.

This concentration of power and resources also influenced recent developments in AI research and have stirred the direction of the field. For instance, current discussions on the ethical implications of AI development were propelled by Big Tech funding, especially after the Cambridge Analytica scandal gave evidence of the large-scale social implications of platform power \cite{Helmond2019}. In terms of technical and scientific developments in the field, Big Tech also increased the number of partnerships with universities and publicly-funded research institutions, the same that were fundamental in the development of the publicly funded infrastructures of ARPANET during the dawn of the internet. However, these relationships between academia and industry transacted by Big Tech are now characterized by power imbalances in favour of these private corporations. For instance, Ochigame argues that, in the “Partnership on AI” initiative, a not-for-profit coalition between major tech companies and universities, the latter has less decision-making power than Big Tech members, contributing to the idea that some ethical and research initiatives are funded primarily for the benefit of the private sector \cite{Ochigame2019}.

Regarding the funding of academic research by Big Tech, Abdalla and Abdalla found that faculty members in the areas of computer science, AI ethics, and AI fairness from four leading R1 universities (MIT, Stanford, Berkeley, and the University of Toronto), who disclosed funding sources, received grants from major technology companies \cite{Abdalla2020}. Counting the percentage of individuals who received Big Tech funding during any moment of their careers, including their doctoral studies, the number increased to 84\% for computer sciences faculty, 88\% for those working on AI, and 97\% for those working on AI ethics \cite{Abdalla2020}. Funding from private entities in academic research raises questions about what type of advances are pursued and funded and what outcomes are privileged over others.

More concretely, the trend seems to be that large models trained with huge datasets that depend on the existing infrastructures of major technology corporations are privileged over other types of research. This was evidenced by the winner of best paper award at the most recent Conference on Neural Information Processing Systems (NeurIPS), one of the major conferences in artificial intelligence, which presented OpenAI’s Generative Pre-trained Transformer 3 (GPT-3) language model \cite{Hao2021}. Large models like GPT-3, however, present environmental, economic, and social concerns. First, because of their requirement of enormous processing power and electricity that, in the case of language processing models, can emit up to 626,000 pounds of carbon dioxide, the equivalent of around five times the average emissions of a car in the United States \cite{Strubell2020}. And, second, because these “state of the art” models are also trained with large amounts of data that privilege quantity over quality of content, paying little attention to social context, and being difficult to audit \cite{Bender2021}. Bender and colleagues also point out that these large models can manipulate the application of their outputs, potentially benefiting the companies behind them in social and economic settings \cite{Bender2021}.

As in the case of the printing press, the initial euphoria of a technology capable of breaking up powers in the status quo shifted the distribution of power to a new equilibrium. The printing press created new centers of power (around states, printers) and gutted the old centers of power (religious institutions).  Datafication and AI initially gave way to relatively frictionless means to communicate and link up with others (the Internet) that was created by the state.  As the shift towards datafication has intensified, however, the corporations have seized power by controlling the currency of power: data. Coupled with largely corporate-fronted advances in AI, the corporations have taken on new roles in defining and directing the lives of billions of people.

\subsection{Communities of Individuals?}

The printing press and the communication revolution it created laid the groundwork for what Anderson calls “imagined communities” in which members of a large geographic area develop a socially-constructed concept of themselves as members of a group, in that context a nation-state \cite{Anderson1983}. The internet and datafication has ushered in a time in which geographic proximity is no longer essential to the ability to create community. This creates opportunities for communities to develop while also providing the opportunity for these communities to coalesce around potentially narrower interests and information creating concern among some about the development of “filter bubbles.”\footnote{ There is some disagreement about whether filter bubbles are an appropriate term or more akin to a term to incite concern about a new technology \cite{Burns2019}, but for our purpose we use the term because it connotes the idea that there is a filtering process (both by individuals and algorithms) that can create an information and interaction bubble for people.}

At the same time, the communities that are created (by conscious choice and by interaction with unknown algorithms) can now involve ever narrower groups of people with similar beliefs, interests and behaviors, because geography is no longer a condition for interaction. In one sense, this allows people to find and identify with the people with whom they share the most similarity and can be very valuable in that regard. For example, this gave activists an edge in online protests, using methods such as doxing and DDoS attacks on targets \cite{wong_e-bandits_2013}. On the other hand, the ubiquity of information combined with datafiction and predictive algorithms that generate internet search results, recommended new purchases, and help introduce our online personas to others can lead to the creation of  “filter bubbles,” which are a “personal ecosystem of information that's been catered by these algorithms” \cite{Parramore2010}.  

The printing press facilitated the opening of communication and knowledge, breaking the stranglehold of the Church on “the word” and letting new social classes rise with the newfound capability to acquire and spread information.  The Internet, in part, has done this as well \cite{deibert_parchment_1997}.  Media companies, which previously held the key to accessing mass communications such as news and entertainment, were subsumed by the competition created by the entry of Internet-based firms \cite{wu_master_2010}. Yet, because of the way that search, advertising, and newsfeed technologies have pivoted to data-intensive processes of personalized advertising and exposure to information, this has resulted in the somewhat discordant result of simultaneously allowing for new sources of information to emerge and driving people into groups that share a common data stream with each other but not one that is broadly shared with others. Thus, there are both more data sources and fewer options at the same time, driven by datafication and AI.  We both know more about the individuals participating in digital technologies and are better able to push them towards their respective interests.

In studying the online media and information environment, Sood and Lelkes discuss how the development of narrowcasting and the replacement of a few, well-known media sources with a multitude of less-known media sources has given consumers greater choice but also allowed them to choose to consume their information from more congenial information sources, which can create a filter bubble of information \cite{Sood2017}. In the context of the current information environment with its thousands of possible websites and sources of information, datafication and predictive algorithms play an indispensable (if unappreciated by many) role in influencing what people see, read, or otherwise consume. The information sources to which people are exposed are necessarily heavily influenced by the personalization algorithms that determines what choices are available to us.

The choice of more congenial media sources could be driven by either a preference for congenial news or the belief that congenial information sources are more trustworthy \cite{Sood2017}. While there is some evidence that in the aggregate people are reasonable arbiters of quality news sources \cite{Pennycook2019}, this does not imply that all individuals are good arbiters or that when presented with an algorithmically chosen list of news stories/sources that we choose wisely. In fact, Luca et al. demonstrate that some individuals actually prefer low-quality “click bait” stories because they believe them to be more trustworthy \cite{Luca2021}. Perceived trust in an information source is a necessary condition for persuasion \cite{Lupia1998} and therefore these sources can still influence behavior and beliefs. 

The combination of preferences for low-quality (but trusted) sources and predictive algorithms may be particularly negative as it pushes people towards inferior information sources. The algorithms are designed to encourage clicks rather than information quality and therefore, if an individual chooses information from a less reliable source and the algorithm then feeds the person more information from similar sources, then they will be increasingly likely to see stories from unreliable, but perceived to be trustworthy, sources. Although people have agency over what stories to click on and what to read, the options presented to them are determined by algorithms that have no concern with the accuracy of the information or its effects. 

Of course the information streams are not fully individualized, rather datafication allows the creation of communities of individuals who share the same media sources (and which may be quite low quality) and who may be geographically distant from one another. Pushing people into communities  and information environments that are persuasive but full of misinformation can have significant real-world implications. For example, Hillary Clinton, as we know, lost the 2016 Presidential election to Donald Trump.  Yet, that was not the end of her importance in that year, at least among some circles.  An online conspiracy theory flourished on platforms such as Reddit, 4Chan, and Facebook, based on the leaked Clinton emails.  The theory put Clinton at the center of a child trafficking ring that was centered in the basement of  Comet Ping Pong, a pizzeria on Connecticut Avenue in Washington, D.C.  Reddit users concluded that “cheese pizza” was code phrase for “child pornography.”  These theories became so widespread, and the allegations so animated that on December 4, 2016, North Carolinian Edgar Maddison Welch walked in with a military-style rifle and handgun and fired several shots.  No one was hurt.  In January 2019, Californian Ryan Jaselskis set Comet Ping Pong on fire.  Although other platforms had shut down Pizzagate by this time, a group on Facebook, “PizzagateUncompromised,” remained alive and well, even after the theory had been debunked by multiple sources, including The New York Times and Snopes.com. The Pizzagate example demonstrates the power of datafication and the creation of communities that turned online interaction into “real-world” social action. While the Pizzagate episode demonstrates a dangerous combination of community and information, it serves as a vivid illustration of how datafication will affect social and political worlds. 

\section{Going Forward}

The datafication of the world portends social, economic, and political changes that are foreshadowed by the upheaval caused by the printing press. These changes will have both normatively good and bad implications, and our focus in this paper is to argue that we think carefully about the changes that datafication will cause; not because we can stop it (or might even want to) but rather because we need to build a new legal and regulatory infrastructure \cite{Hadfield2016}.  Already, rules are shifting as they adapt to the information age \cite{cohen_between_2019}, but our analysis shows that the slow shift as laws catch up to realities may be too leisurely for anticipating the deep social and political changes that are possible.

An important takeaway from our argument is that in thinking about how to address the challenges of datafication from a policy perspective we need to recognize that the changes wrought by it will be no less profound. Therefore, when designing policy responses to manage datafication we need to move away from thinking about datafication as just a faster printing press and instead we need to incorporate technical and social insights.  It is one thing to think about regulating a technology, such as facial recognition, but another thing to truly assess what it means to have data about a person’s face sitting, even regulated, on servers.  What status quo powers are disrupted, or decentralized, and what actors are poised to seize power in the vacuum and create a world to their liking?  Already, Big Tech’s major players have taken advantage of ideological frameworks, a relatively sanguine attitude towards datafication, and a hands-off attitude towards AI to shape the world as they see fit.  The many of us are captive in those platforms without much protection, but that is not the only way.

The way we discuss potential regulatory frameworks about these issues should harken back to the analog world of the printing press.  What rules became possible in a world of print that was impossible in a world of spoken words?  What rules have become impossible in a world of digital data that were possible in the world of print?  A number of basic human rights – freedom of expression, right to privacy/consent – have already come to the fore as problems of portability into the digital era. We simply cannot assume that our rules from twenty years past will apply now or twenty years hence.  Instead, learning from the effects of the printing press, we should look to who holds the power in a digital world, why, and what constitutive changes to expect from the technology.

\begin{acks}
Funded by the Canada Research Chair program and the Schwartz Reisman Institute for Technology and Society. The authors would like to thank the anonymous reviewers for their feedback, and James Long and Josie Greenhill for their valuable comments on earlier versions of this work.
\end{acks}

\bibliographystyle{ACM-Reference-Format}
\bibliography{library.bib}

%%% -*-BibTeX-*-
%%% Do NOT edit. File created by BibTeX with style
%%% ACM-Reference-Format-Journals [18-Jan-2012].

\begin{thebibliography}{96}

%%% ====================================================================
%%% NOTE TO THE USER: you can override these defaults by providing
%%% customized versions of any of these macros before the \bibliography
%%% command.  Each of them MUST provide its own final punctuation,
%%% except for \shownote{}, \showDOI{}, and \showURL{}.  The latter two
%%% do not use final punctuation, in order to avoid confusing it with
%%% the Web address.
%%%
%%% To suppress output of a particular field, define its macro to expand
%%% to an empty string, or better, \unskip, like this:
%%%
%%% \newcommand{\showDOI}[1]{\unskip}   % LaTeX syntax
%%%
%%% \def \showDOI #1{\unskip}           % plain TeX syntax
%%%
%%% ====================================================================

\ifx \showCODEN    \undefined \def \showCODEN     #1{\unskip}     \fi
\ifx \showDOI      \undefined \def \showDOI       #1{#1}\fi
\ifx \showISBNx    \undefined \def \showISBNx     #1{\unskip}     \fi
\ifx \showISBNxiii \undefined \def \showISBNxiii  #1{\unskip}     \fi
\ifx \showISSN     \undefined \def \showISSN      #1{\unskip}     \fi
\ifx \showLCCN     \undefined \def \showLCCN      #1{\unskip}     \fi
\ifx \shownote     \undefined \def \shownote      #1{#1}          \fi
\ifx \showarticletitle \undefined \def \showarticletitle #1{#1}   \fi
\ifx \showURL      \undefined \def \showURL       {\relax}        \fi
% The following commands are used for tagged output and should be
% invisible to TeX
\providecommand\bibfield[2]{#2}
\providecommand\bibinfo[2]{#2}
\providecommand\natexlab[1]{#1}
\providecommand\showeprint[2][]{arXiv:#2}

\bibitem[\protect\citeauthoryear{Abdalla and Abdalla}{Abdalla and
  Abdalla}{2020}]%
        {Abdalla2020}
\bibfield{author}{\bibinfo{person}{Mohamed Abdalla} {and}
  \bibinfo{person}{Moustafa Abdalla}.} \bibinfo{year}{2020}\natexlab{}.
\newblock \showarticletitle{{The Grey Hoodie Project: Big Tobacco, Big Tech,
  and the threat on academic integrity}}.
\newblock \bibinfo{journal}{\emph{arXiv}} (\bibinfo{year}{2020}).
\newblock
\showeprint[arxiv]{2009.13676}
\urldef\tempurl%
\url{http://arxiv.org/abs/2009.13676}
\showURL{%
\tempurl}


\bibitem[\protect\citeauthoryear{Agrawal, Gans, and Goldfarb}{Agrawal
  et~al\mbox{.}}{2018}]%
        {Agrawal2018}
\bibfield{author}{\bibinfo{person}{Jay Agrawal}, \bibinfo{person}{Joshua Gans},
  {and} \bibinfo{person}{Avi Goldfarb}.} \bibinfo{year}{2018}\natexlab{}.
\newblock \bibinfo{booktitle}{\emph{{Prediction Machines: The Simple Economics
  of Artificial Intelligence}}}.
\newblock \bibinfo{publisher}{Harvard Business Review Press},
  \bibinfo{address}{Cambridge, MA}.
\newblock


\bibitem[\protect\citeauthoryear{Anderson}{Anderson}{1983}]%
        {Anderson1983}
\bibfield{author}{\bibinfo{person}{Benedict Anderson}.}
  \bibinfo{year}{1983}\natexlab{}.
\newblock \bibinfo{booktitle}{\emph{{Imagined Communities: Reflections on the
  Origin and Spread of Nationalism}}}.
\newblock \bibinfo{publisher}{Verso}, \bibinfo{address}{London}. 160 pages.
\newblock


\bibitem[\protect\citeauthoryear{Anthes}{Anthes}{2015}]%
        {Anthes2015}
\bibfield{author}{\bibinfo{person}{Gary Anthes}.}
  \bibinfo{year}{2015}\natexlab{}.
\newblock \showarticletitle{{Data brokers are watching you}}.
\newblock \bibinfo{journal}{\emph{Commun. ACM}} \bibinfo{volume}{58},
  \bibinfo{number}{1} (\bibinfo{year}{2015}), \bibinfo{pages}{28--30}.
\newblock


\bibitem[\protect\citeauthoryear{Arora}{Arora}{2019}]%
        {Arora2019}
\bibfield{author}{\bibinfo{person}{Payal Arora}.}
  \bibinfo{year}{2019}\natexlab{}.
\newblock \showarticletitle{{Decolonizing Privacy Studies}}.
\newblock \bibinfo{journal}{\emph{Television {\&} New Media}}
  \bibinfo{volume}{20}, \bibinfo{number}{4} (\bibinfo{date}{may}
  \bibinfo{year}{2019}), \bibinfo{pages}{366--378}.
\newblock


\bibitem[\protect\citeauthoryear{Bender, Gebru, McMillan-Major, and
  Shmitchell}{Bender et~al\mbox{.}}{2021}]%
        {Bender2021}
\bibfield{author}{\bibinfo{person}{Emily~M. Bender}, \bibinfo{person}{Timnit
  Gebru}, \bibinfo{person}{Angelina McMillan-Major}, {and}
  \bibinfo{person}{Shmargaret Shmitchell}.} \bibinfo{year}{2021}\natexlab{}.
\newblock \showarticletitle{{On the Dangers of Stochastic Parrots: Can Language
  Models Be Too Big?}}
\newblock \bibinfo{journal}{\emph{Conference on Fairness, Accountability, and
  Transparency (FAccT '21)}} (\bibinfo{date}{mar} \bibinfo{year}{2021}).
\newblock


\bibitem[\protect\citeauthoryear{Benjamin}{Benjamin}{2019}]%
        {benjamin_race_2019}
\bibfield{author}{\bibinfo{person}{Ruha Benjamin}.}
  \bibinfo{year}{2019}\natexlab{}.
\newblock \bibinfo{booktitle}{\emph{Race after technology: abolitionist tools
  for the {New} {Jim} {Code}}}.
\newblock \bibinfo{publisher}{Polity Press}, \bibinfo{address}{Cambridge, UK}.
\newblock


\bibitem[\protect\citeauthoryear{Birhane}{Birhane}{2020}]%
        {Birhane2020}
\bibfield{author}{\bibinfo{person}{Abeba Birhane}.}
  \bibinfo{year}{2020}\natexlab{}.
\newblock \showarticletitle{{Algorithmic Colonization of Africa}}.
\newblock \bibinfo{journal}{\emph{Scripted}} \bibinfo{volume}{17},
  \bibinfo{number}{2} (\bibinfo{year}{2020}), \bibinfo{pages}{389--409}.
\newblock


\bibitem[\protect\citeauthoryear{Brown}{Brown}{2019}]%
        {Brown2019}
\bibfield{author}{\bibinfo{person}{Lane Brown}.}
  \bibinfo{year}{2019}\natexlab{}.
\newblock \showarticletitle{{There Will Be No Turning Back on Facial
  Recognition}}.
\newblock \bibinfo{journal}{\emph{New York Magazine}} (\bibinfo{date}{nov}
  \bibinfo{year}{2019}).
\newblock
\urldef\tempurl%
\url{https://nymag.com/intelligencer/2019/11/the-future-of-facial-recognition-in-america.html}
\showURL{%
\tempurl}


\bibitem[\protect\citeauthoryear{Buolamwini and Gebru}{Buolamwini and
  Gebru}{2018}]%
        {Buolamwini2018}
\bibfield{author}{\bibinfo{person}{Joy Buolamwini} {and}
  \bibinfo{person}{Timnit Gebru}.} \bibinfo{year}{2018}\natexlab{}.
\newblock \showarticletitle{{Gender Shades: Intersectional Accuracy Disparities
  in Commercial Gender Classification}}.
\newblock \bibinfo{journal}{\emph{Proceedings of the 1st Conference on
  Fairness, Accountability and Transparency}} (\bibinfo{year}{2018}).
\newblock


\bibitem[\protect\citeauthoryear{Burgess}{Burgess}{2019}]%
        {Burgess2019}
\bibfield{author}{\bibinfo{person}{Matt Burgess}.}
  \bibinfo{year}{2019}\natexlab{}.
\newblock \showarticletitle{{The Met Police's facial recognition tests are
  fatally flawed}}.
\newblock \bibinfo{journal}{\emph{WIRED}} (\bibinfo{year}{2019}).
\newblock
\urldef\tempurl%
\url{https://www.wired.co.uk/article/met-police-london-facial-recognition-test}
\showURL{%
\tempurl}


\bibitem[\protect\citeauthoryear{Burns}{Burns}{2019}]%
        {Burns2019}
\bibfield{author}{\bibinfo{person}{Ryan Burns}.}
  \bibinfo{year}{2019}\natexlab{}.
\newblock \showarticletitle{{New Frontiers of Philanthro-capitalism: Digital
  Technologies and Humanitarianism}}.
\newblock \bibinfo{journal}{\emph{Antipode}} \bibinfo{volume}{51},
  \bibinfo{number}{4} (\bibinfo{year}{2019}), \bibinfo{pages}{1101--1122}.
\newblock


\bibitem[\protect\citeauthoryear{Burrell}{Burrell}{2016}]%
        {Burrell2016}
\bibfield{author}{\bibinfo{person}{Jenna Burrell}.}
  \bibinfo{year}{2016}\natexlab{}.
\newblock \showarticletitle{{How the machine ‘thinks': Understanding opacity
  in machine learning algorithms}}.
\newblock \bibinfo{journal}{\emph{Big Data {\&} Society}} \bibinfo{volume}{3},
  \bibinfo{number}{1} (\bibinfo{date}{jan} \bibinfo{year}{2016}).
\newblock


\bibitem[\protect\citeauthoryear{Calo}{Calo}{2017}]%
        {Calo2017}
\bibfield{author}{\bibinfo{person}{Ryan Calo}.}
  \bibinfo{year}{2017}\natexlab{}.
\newblock \showarticletitle{{Artificial Intelligence Policy: A Roadmap}}.
\newblock \bibinfo{journal}{\emph{SSRN Electronic Journal}}
  (\bibinfo{year}{2017}), \bibinfo{pages}{1--28}.
\newblock


\bibitem[\protect\citeauthoryear{Canon}{Canon}{2019}]%
        {Canon2019}
\bibfield{author}{\bibinfo{person}{Gabrielle Canon}.}
  \bibinfo{year}{2019}\natexlab{}.
\newblock \bibinfo{title}{{How Taylor Swift showed us the scary future of
  facial recognition}}.
\newblock
\newblock
\urldef\tempurl%
\url{https://www.theguardian.com/technology/2019/feb/15/how-taylor-swift-showed-us-the-scary-future-of-facial-recognition}
\showURL{%
\tempurl}


\bibitem[\protect\citeauthoryear{Casilli}{Casilli}{2019}]%
        {Casilli2019a}
\bibfield{author}{\bibinfo{person}{Antonio~A. Casilli}.}
  \bibinfo{year}{2019}\natexlab{}.
\newblock \bibinfo{booktitle}{\emph{{En attendant les robots}}}.
\newblock \bibinfo{publisher}{{\'{E}}ditions du Seuil},
  \bibinfo{address}{Paris}. 401 pages.
\newblock


\bibitem[\protect\citeauthoryear{Casilli and Posada}{Casilli and
  Posada}{2019}]%
        {Casilli2019}
\bibfield{author}{\bibinfo{person}{Antonio~A. Casilli} {and}
  \bibinfo{person}{Julian Posada}.} \bibinfo{year}{2019}\natexlab{}.
\newblock \showarticletitle{{The Platformisation of Labor and Society}}.
\newblock In \bibinfo{booktitle}{\emph{Society and the Internet}
  (\bibinfo{edition}{2} ed.)}, \bibfield{editor}{\bibinfo{person}{Mark Graham}
  {and} \bibinfo{person}{William~H. Dutton}} (Eds.). \bibinfo{publisher}{Oxford
  University Press}, \bibinfo{address}{Oxford}.
\newblock


\bibitem[\protect\citeauthoryear{Chokshi}{Chokshi}{2019}]%
        {Chokshi2019}
\bibfield{author}{\bibinfo{person}{Niraj Chokshi}.}
  \bibinfo{year}{2019}\natexlab{}.
\newblock \showarticletitle{{Facial Recognition's Many Controversies, From
  Stadium Surveillance to Racist Software}}.
\newblock \bibinfo{journal}{\emph{The New York Times}} (\bibinfo{year}{2019}).
\newblock
\urldef\tempurl%
\url{https://www.nytimes.com/2019/05/15/business/facial-recognition-software-controversy.html}
\showURL{%
\tempurl}


\bibitem[\protect\citeauthoryear{Chouldechova and Roth}{Chouldechova and
  Roth}{2018}]%
        {Chouldechova2018}
\bibfield{author}{\bibinfo{person}{Alexandra Chouldechova} {and}
  \bibinfo{person}{Aaron Roth}.} \bibinfo{year}{2018}\natexlab{}.
\newblock \showarticletitle{{The frontiers of fairness in machine learning}}.
\newblock \bibinfo{journal}{\emph{arXiv}} (\bibinfo{year}{2018}).
\newblock
\showeprint[arxiv]{1810.08810}
\urldef\tempurl%
\url{https://arxiv.org/abs/1810.08810}
\showURL{%
\tempurl}


\bibitem[\protect\citeauthoryear{Clarke}{Clarke}{1988}]%
        {clarke_information_1988}
\bibfield{author}{\bibinfo{person}{Roger Clarke}.}
  \bibinfo{year}{1988}\natexlab{}.
\newblock \showarticletitle{Information {Technology} and {Dataveillance}}.
\newblock \bibinfo{journal}{\emph{Communication ACM}} \bibinfo{volume}{31},
  \bibinfo{number}{5} (\bibinfo{year}{1988}), \bibinfo{pages}{498--512}.
\newblock


\bibitem[\protect\citeauthoryear{Clarke and Greenleaf}{Clarke and
  Greenleaf}{2017}]%
        {clarke_dataveillance_2017}
\bibfield{author}{\bibinfo{person}{Roger Clarke} {and} \bibinfo{person}{Graham
  Greenleaf}.} \bibinfo{year}{2017}\natexlab{}.
\newblock \showarticletitle{Dataveillance {Regulation}: {A} {Research}
  {Framework}}.
\newblock \bibinfo{journal}{\emph{Journal of Law, Information and Science}}
  \bibinfo{volume}{25}, \bibinfo{number}{1} (\bibinfo{year}{2017}),
  \bibinfo{pages}{104--122}.
\newblock


\bibitem[\protect\citeauthoryear{Cohen}{Cohen}{2019}]%
        {cohen_between_2019}
\bibfield{author}{\bibinfo{person}{Julie~E. Cohen}.}
  \bibinfo{year}{2019}\natexlab{}.
\newblock \bibinfo{booktitle}{\emph{Between truth and power: the legal
  constructions of informational capitalism}}.
\newblock \bibinfo{publisher}{Oxford University Press}, \bibinfo{address}{New
  York, NY}.
\newblock


\bibitem[\protect\citeauthoryear{Corbett-Davies and Goel}{Corbett-Davies and
  Goel}{2018}]%
        {Corbett-Davies2018}
\bibfield{author}{\bibinfo{person}{Sam Corbett-Davies} {and}
  \bibinfo{person}{Sharad Goel}.} \bibinfo{year}{2018}\natexlab{}.
\newblock \showarticletitle{{The measure and mismeasure of fairness: A critical
  review of fair machine learning}}.
\newblock \bibinfo{journal}{\emph{arXiv}} (\bibinfo{year}{2018}).
\newblock
\showeprint[arxiv]{1808.00023}
\urldef\tempurl%
\url{https://arxiv.org/abs/1808.00023}
\showURL{%
\tempurl}


\bibitem[\protect\citeauthoryear{Couldry and Yu}{Couldry and Yu}{2018}]%
        {couldry_deconstructing_2018}
\bibfield{author}{\bibinfo{person}{Nick Couldry} {and} \bibinfo{person}{Jun
  Yu}.} \bibinfo{year}{2018}\natexlab{}.
\newblock \showarticletitle{Deconstructing datafication’s brave new world}.
\newblock \bibinfo{journal}{\emph{New Media \& Society}} \bibinfo{volume}{20},
  \bibinfo{number}{12} (\bibinfo{date}{Dec.} \bibinfo{year}{2018}),
  \bibinfo{pages}{4473--4491}.
\newblock


\bibitem[\protect\citeauthoryear{Deibert}{Deibert}{1997}]%
        {deibert_parchment_1997}
\bibfield{author}{\bibinfo{person}{Ronald Deibert}.}
  \bibinfo{year}{1997}\natexlab{}.
\newblock \bibinfo{booktitle}{\emph{Parchment, printing, and hypermedia:
  communication in world order transformation}}.
\newblock \bibinfo{publisher}{Columbia University Press}, \bibinfo{address}{New
  York}.
\newblock


\bibitem[\protect\citeauthoryear{Deibert}{Deibert}{2020}]%
        {deibert_reset_2020}
\bibfield{author}{\bibinfo{person}{Ronald~J. Deibert}.}
  \bibinfo{year}{2020}\natexlab{}.
\newblock \bibinfo{booktitle}{\emph{Reset: {Reclaiming} the {Internet} for
  {Civil} {Society}}}.
\newblock \bibinfo{publisher}{House of Anansi Press},
  \bibinfo{address}{Toronto, ON}.
\newblock


\bibitem[\protect\citeauthoryear{Denton, Hanna, Amironesei, Smart, Nicole, and
  Scheuerman}{Denton et~al\mbox{.}}{2020}]%
        {Denton2020}
\bibfield{author}{\bibinfo{person}{Emily Denton}, \bibinfo{person}{Alex Hanna},
  \bibinfo{person}{Razvan Amironesei}, \bibinfo{person}{Andrew Smart},
  \bibinfo{person}{Hilary Nicole}, {and} \bibinfo{person}{Morgan~Klaus
  Scheuerman}.} \bibinfo{year}{2020}\natexlab{}.
\newblock \showarticletitle{{Bringing the People Back In: Contesting Benchmark
  Machine Learning Datasets}}.
\newblock \bibinfo{journal}{\emph{arXiv}} (\bibinfo{year}{2020}).
\newblock
\showeprint[arxiv]{2007.07399}
\urldef\tempurl%
\url{http://arxiv.org/abs/2007.07399}
\showURL{%
\tempurl}


\bibitem[\protect\citeauthoryear{Dewar}{Dewar}{1998}]%
        {Dewar1998}
\bibfield{author}{\bibinfo{person}{James Dewar}.}
  \bibinfo{year}{1998}\natexlab{}.
\newblock \bibinfo{booktitle}{\emph{{The Information Age and the Printing
  Press: Looking Backward to See Ahead}}}.
\newblock \bibinfo{publisher}{RAND Corporation}.
\newblock
\urldef\tempurl%
\url{https://www.rand.org/pubs/papers/P8014.html}
\showURL{%
\tempurl}


\bibitem[\protect\citeauthoryear{Domo}{Domo}{2020}]%
        {Domo2020}
\bibfield{author}{\bibinfo{person}{Domo}.} \bibinfo{year}{2020}\natexlab{}.
\newblock \showarticletitle{{Data Never Sleeps 7.0}}.
\newblock \bibinfo{journal}{\emph{Domo}} (\bibinfo{year}{2020}).
\newblock
\urldef\tempurl%
\url{https://www.domo.com/learn/data-never-sleeps-5}
\showURL{%
\tempurl}


\bibitem[\protect\citeauthoryear{Dwork, Hardt, Pitassi, Reingold, and
  Zemel}{Dwork et~al\mbox{.}}{2012}]%
        {Dwork2012}
\bibfield{author}{\bibinfo{person}{Cynthia Dwork}, \bibinfo{person}{Moritz
  Hardt}, \bibinfo{person}{Toniann Pitassi}, \bibinfo{person}{Omer Reingold},
  {and} \bibinfo{person}{Richard Zemel}.} \bibinfo{year}{2012}\natexlab{}.
\newblock \showarticletitle{{Fairness through awareness}}.
\newblock \bibinfo{journal}{\emph{ITCS 2012 - Innovations in Theoretical
  Computer Science Conference}} (\bibinfo{year}{2012}),
  \bibinfo{pages}{214--226}.
\newblock
\showISBNx{9781450311151}
\showeprint[arxiv]{1104.3913}


\bibitem[\protect\citeauthoryear{Eisenstein}{Eisenstein}{1979}]%
        {eisenstein_printing_1979}
\bibfield{author}{\bibinfo{person}{Elizabeth~L. Eisenstein}.}
  \bibinfo{year}{1979}\natexlab{}.
\newblock \bibinfo{booktitle}{\emph{The printing press as an agent of change:
  communications and cultural transformations in early-modern {Europe}.
  {Volumes} {I} and {II}}}.
\newblock \bibinfo{publisher}{Cambridge University Press},
  \bibinfo{address}{Cambridge}.
\newblock


\bibitem[\protect\citeauthoryear{Foroohar}{Foroohar}{2019}]%
        {foroohar2019don}
\bibfield{author}{\bibinfo{person}{Rana Foroohar}.}
  \bibinfo{year}{2019}\natexlab{}.
\newblock \bibinfo{booktitle}{\emph{{Don't Be Evil}}}.
\newblock \bibinfo{publisher}{Currency}, \bibinfo{address}{New York, NY}.
\newblock


\bibitem[\protect\citeauthoryear{Fourcade and Healy}{Fourcade and
  Healy}{2017}]%
        {fourcade_seeing_2017}
\bibfield{author}{\bibinfo{person}{Marion Fourcade} {and}
  \bibinfo{person}{Kieran Healy}.} \bibinfo{year}{2017}\natexlab{}.
\newblock \showarticletitle{Seeing like a market}.
\newblock \bibinfo{journal}{\emph{Socio-Economic Review}}
  (\bibinfo{year}{2017}).
\newblock


\bibitem[\protect\citeauthoryear{Frischmann and Selinger}{Frischmann and
  Selinger}{2018}]%
        {frischmann_re-engineering_2018}
\bibfield{author}{\bibinfo{person}{Brett~M. Frischmann} {and}
  \bibinfo{person}{Evan Selinger}.} \bibinfo{year}{2018}\natexlab{}.
\newblock \bibinfo{booktitle}{\emph{Re-engineering humanity}}.
\newblock \bibinfo{publisher}{Cambridge University Press},
  \bibinfo{address}{Cambridge}.
\newblock


\bibitem[\protect\citeauthoryear{Gillies and Cailliau}{Gillies and
  Cailliau}{2000}]%
        {gillies2000web}
\bibfield{author}{\bibinfo{person}{James~M Gillies} {and}
  \bibinfo{person}{Robert Cailliau}.} \bibinfo{year}{2000}\natexlab{}.
\newblock \bibinfo{booktitle}{\emph{{How the Web was born: The story of the
  World Wide Web}}}.
\newblock \bibinfo{publisher}{Oxford University Press},
  \bibinfo{address}{Oxford}.
\newblock


\bibitem[\protect\citeauthoryear{Hadfield}{Hadfield}{2016}]%
        {Hadfield2016}
\bibfield{author}{\bibinfo{person}{Gillian~K. Hadfield}.}
  \bibinfo{year}{2016}\natexlab{}.
\newblock \bibinfo{booktitle}{\emph{{Rules for a Flat World: Why Humans
  Invented Law and How to Reinvent It for a Complex Global Economy}}}.
\newblock \bibinfo{publisher}{Oxford University Press},
  \bibinfo{address}{Oxford}.
\newblock


\bibitem[\protect\citeauthoryear{Hadfield-Menell, Andrus, and
  Hadfield}{Hadfield-Menell et~al\mbox{.}}{2019}]%
        {Hadfield-Menell2019a}
\bibfield{author}{\bibinfo{person}{Dylan Hadfield-Menell},
  \bibinfo{person}{McKane Andrus}, {and} \bibinfo{person}{Gillian~K.
  Hadfield}.} \bibinfo{year}{2019}\natexlab{}.
\newblock \showarticletitle{{Legible normativity for AI alignment: The value of
  silly rules}}.
\newblock \bibinfo{journal}{\emph{AIES 2019 - Proceedings of the 2019 AAAI/ACM
  Conference on AI, Ethics, and Society}} (\bibinfo{year}{2019}),
  \bibinfo{pages}{115--121}.
\newblock
\showeprint[arxiv]{1811.01267}


\bibitem[\protect\citeauthoryear{Hanna and Park}{Hanna and Park}{2020}]%
        {Hanna2020a}
\bibfield{author}{\bibinfo{person}{Alex Hanna} {and} \bibinfo{person}{Tina~M.
  Park}.} \bibinfo{year}{2020}\natexlab{}.
\newblock \showarticletitle{{Against Scale: Provocations and Resistances to
  Scale Thinking}}.
\newblock \bibinfo{journal}{\emph{arXiv}} (\bibinfo{year}{2020}).
\newblock
\showeprint[arxiv]{2010.08850}
\urldef\tempurl%
\url{http://arxiv.org/abs/2010.08850}
\showURL{%
\tempurl}


\bibitem[\protect\citeauthoryear{Hao}{Hao}{2020}]%
        {Hao2020a}
\bibfield{author}{\bibinfo{person}{Karen Hao}.}
  \bibinfo{year}{2020}\natexlab{}.
\newblock \showarticletitle{{In 2020, let's stop AI ethics-washing and actually
  do something}}.
\newblock \bibinfo{journal}{\emph{MIT Technology Review}}
  (\bibinfo{year}{2020}).
\newblock
\urldef\tempurl%
\url{https://www.technologyreview.com/2019/12/27/57/ai-ethics-washing-time-to-act/}
\showURL{%
\tempurl}


\bibitem[\protect\citeauthoryear{Hao}{Hao}{2021}]%
        {Hao2021}
\bibfield{author}{\bibinfo{person}{Karen Hao}.}
  \bibinfo{year}{2021}\natexlab{}.
\newblock \showarticletitle{{Five ways to make AI a greater force for good in
  2021}}.
\newblock \bibinfo{journal}{\emph{MIT Technology Review}}
  (\bibinfo{year}{2021}).
\newblock
\urldef\tempurl%
\url{https://www.technologyreview.com/2021/01/08/1015907/ai-force-for-good-in-2021/}
\showURL{%
\tempurl}


\bibitem[\protect\citeauthoryear{Harvey}{Harvey}{2007}]%
        {harvey2007brief}
\bibfield{author}{\bibinfo{person}{David Harvey}.}
  \bibinfo{year}{2007}\natexlab{}.
\newblock \bibinfo{booktitle}{\emph{{A brief history of neoliberalism}}}.
\newblock \bibinfo{publisher}{Oxford University Press, USA}.
\newblock


\bibitem[\protect\citeauthoryear{Helmond, Nieborg, and van~der Vlist}{Helmond
  et~al\mbox{.}}{2019}]%
        {Helmond2019}
\bibfield{author}{\bibinfo{person}{Anne Helmond}, \bibinfo{person}{David~B.
  Nieborg}, {and} \bibinfo{person}{Fernando~N. van~der Vlist}.}
  \bibinfo{year}{2019}\natexlab{}.
\newblock \showarticletitle{{Facebook's evolution: development of a
  platform-as-infrastructure}}.
\newblock \bibinfo{journal}{\emph{Internet Histories}} \bibinfo{volume}{3},
  \bibinfo{number}{2} (\bibinfo{year}{2019}), \bibinfo{pages}{123--146}.
\newblock


\bibitem[\protect\citeauthoryear{Hildebrandt}{Hildebrandt}{2019}]%
        {Hildebrandt2019}
\bibfield{author}{\bibinfo{person}{Mireille Hildebrandt}.}
  \bibinfo{year}{2019}\natexlab{}.
\newblock \showarticletitle{{Privacy as Protection of the Incomputable Self :
  From Agnostic to Agonistic Machine Learning}}.
\newblock \bibinfo{journal}{\emph{Theoretical Inquiries in Law}}
  \bibinfo{volume}{20}, \bibinfo{number}{83} (\bibinfo{year}{2019}),
  \bibinfo{pages}{83--121}.
\newblock


\bibitem[\protect\citeauthoryear{Hill}{Hill}{2020a}]%
        {Hill2020c}
\bibfield{author}{\bibinfo{person}{Kashmir Hill}.}
  \bibinfo{year}{2020}\natexlab{a}.
\newblock \showarticletitle{{Flawed Facial Recognition Leads To Arrest and Jail
  for New Jersey Man}}.
\newblock \bibinfo{journal}{\emph{The New York Times}} (\bibinfo{year}{2020}).
\newblock
\urldef\tempurl%
\url{https://www.nytimes.com/2020/12/29/technology/facial-recognition-misidentify-jail.html}
\showURL{%
\tempurl}


\bibitem[\protect\citeauthoryear{Hill}{Hill}{2020b}]%
        {Hill2020b}
\bibfield{author}{\bibinfo{person}{Kashmir Hill}.}
  \bibinfo{year}{2020}\natexlab{b}.
\newblock \showarticletitle{{The Secretive Company That Might End Privacy as We
  Know It}}.
\newblock \bibinfo{journal}{\emph{The New York Times}} (\bibinfo{year}{2020}).
\newblock
\urldef\tempurl%
\url{https://www.nytimes.com/2020/01/18/technology/clearview-privacy-facial-recognition.html}
\showURL{%
\tempurl}


\bibitem[\protect\citeauthoryear{Ichihashi}{Ichihashi}{2019}]%
        {Ichihashi2019}
\bibfield{author}{\bibinfo{person}{Shota Ichihashi}.}
  \bibinfo{year}{2019}\natexlab{}.
\newblock \showarticletitle{{Non-competing Data Intermediaries}}.
\newblock \bibinfo{journal}{\emph{Bank of Canada Working Paper}}
  (\bibinfo{year}{2019}), \bibinfo{pages}{1--41}.
\newblock


\bibitem[\protect\citeauthoryear{Jo and Gebru}{Jo and Gebru}{2020}]%
        {Jo2020}
\bibfield{author}{\bibinfo{person}{Eun~Seo Jo} {and} \bibinfo{person}{Timnit
  Gebru}.} \bibinfo{year}{2020}\natexlab{}.
\newblock \showarticletitle{{Lessons from archives: Strategies for collecting
  sociocultural data in machine learning}}.
\newblock \bibinfo{journal}{\emph{FAT* 2020 - Proceedings of the 2020
  Conference on Fairness, Accountability, and Transparency}}
  (\bibinfo{year}{2020}), \bibinfo{pages}{306--316}.
\newblock
\showISBNx{9781450369367}
\showeprint[arxiv]{1912.10389}


\bibitem[\protect\citeauthoryear{Jobin, Ienca, and Vayena}{Jobin
  et~al\mbox{.}}{2019}]%
        {Jobin2019}
\bibfield{author}{\bibinfo{person}{Anna Jobin}, \bibinfo{person}{Marcello
  Ienca}, {and} \bibinfo{person}{Effy Vayena}.}
  \bibinfo{year}{2019}\natexlab{}.
\newblock \showarticletitle{{The global landscape of AI ethics guidelines}}.
\newblock \bibinfo{journal}{\emph{Nature Machine Intelligence}}
  \bibinfo{volume}{1}, \bibinfo{number}{9} (\bibinfo{year}{2019}),
  \bibinfo{pages}{389--399}.
\newblock


\bibitem[\protect\citeauthoryear{Jones and Tonetti}{Jones and Tonetti}{2020}]%
        {Jones2020}
\bibfield{author}{\bibinfo{person}{Charles~I. Jones} {and}
  \bibinfo{person}{Christopher Tonetti}.} \bibinfo{year}{2020}\natexlab{}.
\newblock \showarticletitle{{Nonrivalry and the Economics of Data}}.
\newblock \bibinfo{journal}{\emph{American Economic Review}}
  \bibinfo{volume}{110}, \bibinfo{number}{9} (\bibinfo{year}{2020}),
  \bibinfo{pages}{2819--2858}.
\newblock


\bibitem[\protect\citeauthoryear{Kalluri}{Kalluri}{2020}]%
        {Kalluri2020}
\bibfield{author}{\bibinfo{person}{Pratyusha Kalluri}.}
  \bibinfo{year}{2020}\natexlab{}.
\newblock \showarticletitle{{Don't ask if artificial intelligence is good or
  fair, ask how it shifts power}}.
\newblock \bibinfo{journal}{\emph{Nature}} \bibinfo{volume}{583},
  \bibinfo{number}{7815} (\bibinfo{date}{jul} \bibinfo{year}{2020}),
  \bibinfo{pages}{169--169}.
\newblock


\bibitem[\protect\citeauthoryear{Karppi}{Karppi}{2018}]%
        {karppi_disconnect_2018}
\bibfield{author}{\bibinfo{person}{Tero Karppi}.}
  \bibinfo{year}{2018}\natexlab{}.
\newblock \bibinfo{booktitle}{\emph{Disconnect: {Facebook}'s affective bonds}}.
\newblock \bibinfo{publisher}{University of Minnesota Press},
  \bibinfo{address}{Minneapolis}.
\newblock


\bibitem[\protect\citeauthoryear{Kind}{Kind}{2020}]%
        {Kind2020}
\bibfield{author}{\bibinfo{person}{Carly Kind}.}
  \bibinfo{year}{2020}\natexlab{}.
\newblock \showarticletitle{{The term ‘ethical AI' is finally starting to
  mean something}}.
\newblock \bibinfo{journal}{\emph{Venture Beat}} (\bibinfo{date}{aug}
  \bibinfo{year}{2020}).
\newblock
\urldef\tempurl%
\url{https://venturebeat.com/2020/08/23/the-term-ethical-ai-is-finally-starting-to-mean-something/}
\showURL{%
\tempurl}


\bibitem[\protect\citeauthoryear{Kitchin}{Kitchin}{2017}]%
        {kitchin_thinking_2017}
\bibfield{author}{\bibinfo{person}{Rob Kitchin}.}
  \bibinfo{year}{2017}\natexlab{}.
\newblock \showarticletitle{Thinking critically about and researching
  algorithms}.
\newblock \bibinfo{journal}{\emph{Information, Communication \& Society}}
  \bibinfo{volume}{20}, \bibinfo{number}{1} (\bibinfo{date}{Jan.}
  \bibinfo{year}{2017}), \bibinfo{pages}{14--29}.
\newblock


\bibitem[\protect\citeauthoryear{Levy}{Levy}{2013}]%
        {levy_relational_2013}
\bibfield{author}{\bibinfo{person}{Karen Levy}.}
  \bibinfo{year}{2013}\natexlab{}.
\newblock \showarticletitle{Relational {Big} {Data}}.
\newblock \bibinfo{journal}{\emph{{SSRN} {Scholarly} {Paper}}}
  (\bibinfo{date}{Sept.} \bibinfo{year}{2013}).
\newblock
\urldef\tempurl%
\url{https://papers.ssrn.com/abstract=2832921}
\showURL{%
\tempurl}


\bibitem[\protect\citeauthoryear{Lopatto}{Lopatto}{2020}]%
        {Lopatto2020}
\bibfield{author}{\bibinfo{person}{Elizabeth Lopatto}.}
  \bibinfo{year}{2020}\natexlab{}.
\newblock \showarticletitle{{Clearview AI CEO says ‘over 2,400 police
  agencies' are using its facial recognition software}}.
\newblock \bibinfo{journal}{\emph{The Verge}} (\bibinfo{year}{2020}).
\newblock
\urldef\tempurl%
\url{https://www.theverge.com/2020/8/26/21402978/clearview-ai-ceo-interview-2400-police-agencies-facial-recognition}
\showURL{%
\tempurl}


\bibitem[\protect\citeauthoryear{Luca, Munger, Nagler, and Tucker}{Luca
  et~al\mbox{.}}{2021}]%
        {Luca2021}
\bibfield{author}{\bibinfo{person}{Mario Luca}, \bibinfo{person}{Kevin Munger},
  \bibinfo{person}{Jonathan Nagler}, {and} \bibinfo{person}{Joshua~A Tucker}.}
  \bibinfo{year}{2021}\natexlab{}.
\newblock \showarticletitle{{You Won't Believe Our Results! But They Might:
  Heterogeneity in Beliefs About the Accuracy of Online Media}}.
\newblock \bibinfo{journal}{\emph{Journal of Experimental Political Science}}
  (\bibinfo{date}{jan} \bibinfo{year}{2021}), \bibinfo{pages}{1--11}.
\newblock


\bibitem[\protect\citeauthoryear{Lupia and McCubbins}{Lupia and
  McCubbins}{1998}]%
        {Lupia1998}
\bibfield{author}{\bibinfo{person}{Arthur Lupia} {and}
  \bibinfo{person}{Mathew~D. McCubbins}.} \bibinfo{year}{1998}\natexlab{}.
\newblock \bibinfo{booktitle}{\emph{{The Democratic Dilemma}}}.
\newblock \bibinfo{publisher}{Cambridge University Press},
  \bibinfo{address}{Cambridge}.
\newblock


\bibitem[\protect\citeauthoryear{Madden}{Madden}{2017}]%
        {Madden2017}
\bibfield{author}{\bibinfo{person}{Mary Madden}.}
  \bibinfo{year}{2017}\natexlab{}.
\newblock \bibinfo{booktitle}{\emph{{Privacy, Security, and Digital
  Inequality}}}.
\newblock Number September 27. \bibinfo{publisher}{Data {\&} Society},
  \bibinfo{address}{New York, NY}. 125 pages.
\newblock


\bibitem[\protect\citeauthoryear{Marczak, Scott-Railton, Senft, Poetranto, and
  McKune}{Marczak et~al\mbox{.}}{2015}]%
        {Marczak2015}
\bibfield{author}{\bibinfo{person}{Bill Marczak}, \bibinfo{person}{John
  Scott-Railton}, \bibinfo{person}{Adam Senft}, \bibinfo{person}{Irene
  Poetranto}, {and} \bibinfo{person}{Sarah McKune}.}
  \bibinfo{year}{2015}\natexlab{}.
\newblock \showarticletitle{{Pay No Attention to the Server Behind the Proxy}}.
\newblock \bibinfo{journal}{\emph{Citizen Lab}}  \bibinfo{volume}{15}
  (\bibinfo{date}{oct} \bibinfo{year}{2015}).
\newblock


\bibitem[\protect\citeauthoryear{Mayer-Sch{\"{o}}nberger and
  Cukier}{Mayer-Sch{\"{o}}nberger and Cukier}{2014}]%
        {Mayer-Schonberger2014}
\bibfield{author}{\bibinfo{person}{Viktor Mayer-Sch{\"{o}}nberger} {and}
  \bibinfo{person}{Kenneth Cukier}.} \bibinfo{year}{2014}\natexlab{}.
\newblock \bibinfo{booktitle}{\emph{{Big Data: A Revolution That Will Transform
  How We Live, Work, and Think}}}.
\newblock \bibinfo{publisher}{Eamon Dolan Mariner Books},
  \bibinfo{address}{Boston, MA}. 272 pages.
\newblock


\bibitem[\protect\citeauthoryear{Mejias and Couldry}{Mejias and
  Couldry}{2019}]%
        {Mejias2019}
\bibfield{author}{\bibinfo{person}{Ulises~A. Mejias} {and}
  \bibinfo{person}{Nick Couldry}.} \bibinfo{year}{2019}\natexlab{}.
\newblock \showarticletitle{{Datafication}}.
\newblock \bibinfo{journal}{\emph{Internet Policy Review}} \bibinfo{volume}{8},
  \bibinfo{number}{4} (\bibinfo{year}{2019}), \bibinfo{pages}{1--10}.
\newblock


\bibitem[\protect\citeauthoryear{Miceli, Schuessler, and Yang}{Miceli
  et~al\mbox{.}}{2020}]%
        {Miceli2020}
\bibfield{author}{\bibinfo{person}{Milagros Miceli}, \bibinfo{person}{Martin
  Schuessler}, {and} \bibinfo{person}{Tianling Yang}.}
  \bibinfo{year}{2020}\natexlab{}.
\newblock \showarticletitle{{Between Subjectivity and Imposition}}.
\newblock \bibinfo{journal}{\emph{Proceedings of the ACM on Human-Computer
  Interaction}} \bibinfo{volume}{4}, \bibinfo{number}{CSCW2}
  (\bibinfo{year}{2020}), \bibinfo{pages}{1--25}.
\newblock
\showeprint[arxiv]{2007.14886}


\bibitem[\protect\citeauthoryear{Mohamed, Png, and Isaac}{Mohamed
  et~al\mbox{.}}{2020}]%
        {Mohamed2020}
\bibfield{author}{\bibinfo{person}{Shakir Mohamed},
  \bibinfo{person}{Marie-Therese Png}, {and} \bibinfo{person}{William Isaac}.}
  \bibinfo{year}{2020}\natexlab{}.
\newblock \showarticletitle{{Decolonial AI: Decolonial Theory as Sociotechnical
  Foresight in Artificial Intelligence}}.
\newblock \bibinfo{journal}{\emph{Philosophy {\&} Technology}}
  \bibinfo{volume}{33}, \bibinfo{number}{4} (\bibinfo{date}{dec}
  \bibinfo{year}{2020}), \bibinfo{pages}{659--684}.
\newblock
\showeprint[arxiv]{2007.04068}


\bibitem[\protect\citeauthoryear{Newlands}{Newlands}{2021}]%
        {Newlands2021}
\bibfield{author}{\bibinfo{person}{Gemma Newlands}.}
  \bibinfo{year}{2021}\natexlab{}.
\newblock \showarticletitle{{Far from Effortless: Invisible Labour in
  AI-as-a-Service}}.
\newblock \bibinfo{journal}{\emph{Working Paper}} (\bibinfo{year}{2021}).
\newblock


\bibitem[\protect\citeauthoryear{Nieborg and Poell}{Nieborg and Poell}{2018}]%
        {Nieborg2018e}
\bibfield{author}{\bibinfo{person}{David~B. Nieborg} {and}
  \bibinfo{person}{Thomas Poell}.} \bibinfo{year}{2018}\natexlab{}.
\newblock \showarticletitle{{The platformization of cultural production:
  Theorizing the contingent cultural commodity}}.
\newblock \bibinfo{journal}{\emph{New Media {\&} Society}}
  \bibinfo{volume}{20}, \bibinfo{number}{11} (\bibinfo{date}{nov}
  \bibinfo{year}{2018}), \bibinfo{pages}{4275--4292}.
\newblock


\bibitem[\protect\citeauthoryear{Noble}{Noble}{2018}]%
        {noble_algorithms_2018}
\bibfield{author}{\bibinfo{person}{Safiya~Umoja Noble}.}
  \bibinfo{year}{2018}\natexlab{}.
\newblock \bibinfo{booktitle}{\emph{Algorithms of {Oppression}. {How} {Search}
  {Engines} {Reinforce} {Racism}}}.
\newblock \bibinfo{publisher}{NYU Press}, \bibinfo{address}{New York, NY}.
\newblock


\bibitem[\protect\citeauthoryear{Nourbakhsh and Keating}{Nourbakhsh and
  Keating}{2020}]%
        {nourbakhsh_ai_2020}
\bibfield{author}{\bibinfo{person}{Illah~Reza Nourbakhsh} {and}
  \bibinfo{person}{Jennifer Keating}.} \bibinfo{year}{2020}\natexlab{}.
\newblock \bibinfo{booktitle}{\emph{{AI} \& humanity}}.
\newblock \bibinfo{publisher}{The MIT Press}, \bibinfo{address}{Cambridge,
  Massachusetts}.
\newblock


\bibitem[\protect\citeauthoryear{Ochigame}{Ochigame}{2019}]%
        {Ochigame2019}
\bibfield{author}{\bibinfo{person}{Rodrigo Ochigame}.}
  \bibinfo{year}{2019}\natexlab{}.
\newblock \showarticletitle{{The invention of “Ethical AI”: How Big Tech
  Manipulates Academia to Avoid Regulation}}.
\newblock \bibinfo{journal}{\emph{The Intercept}} (\bibinfo{year}{2019}).
\newblock
\urldef\tempurl%
\url{https://theintercept.com/2019/12/20/mit-ethical-ai-artificial-intelligence/}
\showURL{%
\tempurl}


\bibitem[\protect\citeauthoryear{O'Neil}{O'Neil}{2017}]%
        {oneil_weapons_2017}
\bibfield{author}{\bibinfo{person}{Cathy O'Neil}.}
  \bibinfo{year}{2017}\natexlab{}.
\newblock \bibinfo{booktitle}{\emph{Weapons of {Math} {Destruction}: {How}
  {Big} {Data} {Increases} {Inequality} and {Threatens} {Democracy}}}.
\newblock \bibinfo{publisher}{Broadway Books}, \bibinfo{address}{Portland, OR}.
\newblock


\bibitem[\protect\citeauthoryear{Parramore}{Parramore}{2010}]%
        {Parramore2010}
\bibfield{author}{\bibinfo{person}{Lynn Parramore}.}
  \bibinfo{year}{2010}\natexlab{}.
\newblock \showarticletitle{{The Filter Bubble}}.
\newblock \bibinfo{journal}{\emph{The Atlantic}} (\bibinfo{year}{2010}).
\newblock
\urldef\tempurl%
\url{https://www.theatlantic.com/daily-dish/archive/2010/10/the-filter-bubble/181427/}
\showURL{%
\tempurl}


\bibitem[\protect\citeauthoryear{Pasquale}{Pasquale}{2015}]%
        {Pasquale2015b}
\bibfield{author}{\bibinfo{person}{Frank Pasquale}.}
  \bibinfo{year}{2015}\natexlab{}.
\newblock \bibinfo{booktitle}{\emph{{The Black Box Society: The Secret
  Algorithms That Control Money and Information}}}.
\newblock \bibinfo{publisher}{Harvard University Press},
  \bibinfo{address}{Cambridge, MA}. 320 pages.
\newblock


\bibitem[\protect\citeauthoryear{Paullada, Raji, Bender, Denton, and
  Hanna}{Paullada et~al\mbox{.}}{2020}]%
        {Paullada2020}
\bibfield{author}{\bibinfo{person}{Amandalynne Paullada},
  \bibinfo{person}{Inioluwa~Deborah Raji}, \bibinfo{person}{Emily~M. Bender},
  \bibinfo{person}{Emily Denton}, {and} \bibinfo{person}{Alex Hanna}.}
  \bibinfo{year}{2020}\natexlab{}.
\newblock \showarticletitle{{Data and its (dis)contents: A survey of dataset
  development and use in machine learning research}}.
\newblock \bibinfo{journal}{\emph{NeurIPS 2020 Workshop: ML Retrospectives,
  Surveys {\&} Meta-analyses (ML-RSA)}} (\bibinfo{year}{2020}).
\newblock
\showeprint[arxiv]{2012.05345}


\bibitem[\protect\citeauthoryear{Pennycook and Rand}{Pennycook and
  Rand}{2019}]%
        {Pennycook2019}
\bibfield{author}{\bibinfo{person}{Gordon Pennycook} {and}
  \bibinfo{person}{David~G. Rand}.} \bibinfo{year}{2019}\natexlab{}.
\newblock \showarticletitle{{Fighting misinformation on social media using
  crowdsourced judgments of news source quality}}.
\newblock \bibinfo{journal}{\emph{Proceedings of the National Academy of
  Sciences}} \bibinfo{volume}{116}, \bibinfo{number}{7} (\bibinfo{date}{feb}
  \bibinfo{year}{2019}), \bibinfo{pages}{2521--2526}.
\newblock


\bibitem[\protect\citeauthoryear{Poell, Nieborg, and van Dijck}{Poell
  et~al\mbox{.}}{2019}]%
        {Poell2019}
\bibfield{author}{\bibinfo{person}{Thomas Poell}, \bibinfo{person}{David~B.
  Nieborg}, {and} \bibinfo{person}{Jos{\'{e}} van Dijck}.}
  \bibinfo{year}{2019}\natexlab{}.
\newblock \showarticletitle{{Platformisation}}.
\newblock \bibinfo{journal}{\emph{Internet Policy Review}} \bibinfo{volume}{8},
  \bibinfo{number}{4} (\bibinfo{year}{2019}).
\newblock


\bibitem[\protect\citeauthoryear{Posada}{Posada}{2020}]%
        {Posada2020a}
\bibfield{author}{\bibinfo{person}{Julian Posada}.}
  \bibinfo{year}{2020}\natexlab{}.
\newblock \showarticletitle{{The Future of Work Is Here: Toward a Comprehensive
  Approach to Artificial Intelligence and Labour}}.
\newblock \bibinfo{journal}{\emph{Ethics in Context}} (\bibinfo{year}{2020}).
\newblock


\bibitem[\protect\citeauthoryear{Raley}{Raley}{2013}]%
        {raley_dataveillance_2013}
\bibfield{author}{\bibinfo{person}{Rita Raley}.}
  \bibinfo{year}{2013}\natexlab{}.
\newblock \showarticletitle{Dataveillance and counterveillance}.
\newblock In \bibinfo{booktitle}{\emph{Raw {Data} {Is} an {Oxymoron}}},
  \bibfield{editor}{\bibinfo{person}{Lisa Gitelman}} (Ed.).
  \bibinfo{publisher}{MIT Press}, \bibinfo{address}{Cambridge, MA},
  \bibinfo{pages}{121--46}.
\newblock


\bibitem[\protect\citeauthoryear{Raviv}{Raviv}{2020}]%
        {Raviv2020}
\bibfield{author}{\bibinfo{person}{Shaun Raviv}.}
  \bibinfo{year}{2020}\natexlab{}.
\newblock \showarticletitle{{The Secret History of Facial Recognition Shaun}}.
\newblock \bibinfo{journal}{\emph{Wired}} (\bibinfo{year}{2020}).
\newblock
\urldef\tempurl%
\url{https://www.wired.com/story/secret-history-facial-recognition/}
\showURL{%
\tempurl}


\bibitem[\protect\citeauthoryear{{Rold{\'{a}}n Vera}}{{Rold{\'{a}}n
  Vera}}{2013}]%
        {RoldanVera2013}
\bibfield{author}{\bibinfo{person}{Eugenia {Rold{\'{a}}n Vera}}.}
  \bibinfo{year}{2013}\natexlab{}.
\newblock \showarticletitle{{The History of the Book in Latin America
  (including Incas and Aztecs)}}.
\newblock In \bibinfo{booktitle}{\emph{The Book. A Global History}},
  \bibfield{editor}{\bibinfo{person}{Michael~F. Suarez} {and}
  \bibinfo{person}{H.R. Woudhuysen}} (Eds.). \bibinfo{publisher}{Oxford
  University Press}, \bibinfo{address}{Oxford}, \bibinfo{pages}{656--670}.
\newblock


\bibitem[\protect\citeauthoryear{Ruud}{Ruud}{1980}]%
        {Ruud1980}
\bibfield{author}{\bibinfo{person}{Charles~A Ruud}.}
  \bibinfo{year}{1980}\natexlab{}.
\newblock \showarticletitle{{The Printing Press as an Agent of Change}}.
\newblock \bibinfo{journal}{\emph{Imago Mundi}} \bibinfo{volume}{32},
  \bibinfo{number}{1} (\bibinfo{year}{1980}), \bibinfo{pages}{95--97}.
\newblock


\bibitem[\protect\citeauthoryear{Sadowski}{Sadowski}{2019}]%
        {sadowski_when_2019}
\bibfield{author}{\bibinfo{person}{Jathan Sadowski}.}
  \bibinfo{year}{2019}\natexlab{}.
\newblock \showarticletitle{When data is capital: {Datafication}, accumulation,
  and extraction}.
\newblock \bibinfo{journal}{\emph{Big Data \& Society}} \bibinfo{volume}{6},
  \bibinfo{number}{1} (\bibinfo{date}{Jan.} \bibinfo{year}{2019}).
\newblock


\bibitem[\protect\citeauthoryear{Sinha}{Sinha}{2018}]%
        {sinha_real-property_2018}
\bibfield{author}{\bibinfo{person}{G.~Alex Sinha}.}
  \bibinfo{year}{2018}\natexlab{}.
\newblock \showarticletitle{A {Real}-{Property} {Model} of {Privacy}}.
\newblock \bibinfo{journal}{\emph{DePaul Law Review}} \bibinfo{volume}{68},
  \bibinfo{number}{3} (\bibinfo{year}{2018}), \bibinfo{pages}{567--614}.
\newblock


\bibitem[\protect\citeauthoryear{Smyrnaios}{Smyrnaios}{2017}]%
        {Smyrnaios2017}
\bibfield{author}{\bibinfo{person}{Nikos Smyrnaios}.}
  \bibinfo{year}{2017}\natexlab{}.
\newblock \bibinfo{booktitle}{\emph{{Les GAFAM contre internet : une
  {\'{e}}conomie politique du num{\'{e}}rique}}}.
\newblock \bibinfo{publisher}{InaGlobal}, \bibinfo{address}{Paris}.
\newblock


\bibitem[\protect\citeauthoryear{Sood and Lelkes}{Sood and Lelkes}{2018}]%
        {Sood2017}
\bibfield{author}{\bibinfo{person}{Gaurav Sood} {and} \bibinfo{person}{Yphtach
  Lelkes}.} \bibinfo{year}{2018}\natexlab{}.
\newblock \showarticletitle{{Don't Expose Yourself: Discretionary Exposure to
  Political Information}}.
\newblock In \bibinfo{booktitle}{\emph{Oxford Research Encyclopedia of
  Politics}}. \bibinfo{publisher}{Oxford University Press},
  \bibinfo{pages}{1--49}.
\newblock


\bibitem[\protect\citeauthoryear{Strubell, Ganesh, and McCallum}{Strubell
  et~al\mbox{.}}{2020}]%
        {Strubell2020}
\bibfield{author}{\bibinfo{person}{Emma Strubell}, \bibinfo{person}{Ananya
  Ganesh}, {and} \bibinfo{person}{Andrew McCallum}.}
  \bibinfo{year}{2020}\natexlab{}.
\newblock \showarticletitle{{Energy and policy considerations for deep learning
  in NLP}}.
\newblock \bibinfo{journal}{\emph{ACL 2019 - 57th Annual Meeting of the
  Association for Computational Linguistics, Proceedings of the Conference}}
  (\bibinfo{year}{2020}), \bibinfo{pages}{3645--3650}.
\newblock
\showeprint[arxiv]{1906.02243}


\bibitem[\protect\citeauthoryear{Thompson}{Thompson}{1995}]%
        {thompson_media_1995}
\bibfield{author}{\bibinfo{person}{John~B. Thompson}.}
  \bibinfo{year}{1995}\natexlab{}.
\newblock \bibinfo{booktitle}{\emph{The media and modernity: a social theory of
  the media}}.
\newblock \bibinfo{publisher}{Stanford University Press},
  \bibinfo{address}{Stanford, CA}.
\newblock


\bibitem[\protect\citeauthoryear{Torralba and Efros}{Torralba and
  Efros}{2011}]%
        {Torralba2011}
\bibfield{author}{\bibinfo{person}{Antonio Torralba} {and}
  \bibinfo{person}{Alexei~A. Efros}.} \bibinfo{year}{2011}\natexlab{}.
\newblock \showarticletitle{{Unbiased look at dataset bias}}. In
  \bibinfo{booktitle}{\emph{CVPR 2011}}. \bibinfo{publisher}{IEEE},
  \bibinfo{pages}{1521--1528}.
\newblock


\bibitem[\protect\citeauthoryear{Tsing}{Tsing}{2012}]%
        {Tsing2012}
\bibfield{author}{\bibinfo{person}{A.~L. Tsing}.}
  \bibinfo{year}{2012}\natexlab{}.
\newblock \showarticletitle{{ON NONSCALABILITY: The Living World Is Not
  Amenable to Precision-Nested Scales}}.
\newblock \bibinfo{journal}{\emph{Common Knowledge}} \bibinfo{volume}{18},
  \bibinfo{number}{3} (\bibinfo{date}{oct} \bibinfo{year}{2012}),
  \bibinfo{pages}{505--524}.
\newblock


\bibitem[\protect\citeauthoryear{van Dijck}{van Dijck}{2014}]%
        {van_dijck_datafication_2014}
\bibfield{author}{\bibinfo{person}{Jose van Dijck}.}
  \bibinfo{year}{2014}\natexlab{}.
\newblock \showarticletitle{Datafication, dataism and dataveillance: {Big}
  {Data} between scientific paradigm and ideology}.
\newblock \bibinfo{journal}{\emph{Surveillance \& Society}}
  \bibinfo{volume}{12}, \bibinfo{number}{2} (\bibinfo{date}{May}
  \bibinfo{year}{2014}), \bibinfo{pages}{197--208}.
\newblock


\bibitem[\protect\citeauthoryear{Varian}{Varian}{2019}]%
        {Varian2019}
\bibfield{author}{\bibinfo{person}{Hal Varian}.}
  \bibinfo{year}{2019}\natexlab{}.
\newblock \showarticletitle{{Artificial Intelligence, Economics, and Industrial
  Organization}}.
\newblock In \bibinfo{booktitle}{\emph{The Economics of Artificial
  Intelligence: An Agenda}}, \bibfield{editor}{\bibinfo{person}{Ajay Agrawal},
  \bibinfo{person}{Joshua Gans}, {and} \bibinfo{person}{Avi Goldfarb}} (Eds.).
  \bibinfo{publisher}{University of Chicago Press}, \bibinfo{address}{Chicago,
  IL}, \bibinfo{pages}{399--419}.
\newblock


\bibitem[\protect\citeauthoryear{Véliz}{Véliz}{2020}]%
        {veliz_privacy_2020}
\bibfield{author}{\bibinfo{person}{Carissa Véliz}.}
  \bibinfo{year}{2020}\natexlab{}.
\newblock \bibinfo{booktitle}{\emph{Privacy is {Power}: {Why} and how you
  should take back control of your data}}.
\newblock \bibinfo{publisher}{Bantam Press}, \bibinfo{address}{New York}.
\newblock


\bibitem[\protect\citeauthoryear{Webb}{Webb}{2019}]%
        {webb_big_2019}
\bibfield{author}{\bibinfo{person}{Amy Webb}.} \bibinfo{year}{2019}\natexlab{}.
\newblock \bibinfo{booktitle}{\emph{The {Big} {Nine}: how the tech titans and
  their thinking machines could warp humanity}}.
\newblock \bibinfo{publisher}{PublicAffairs}, \bibinfo{address}{New York}.
\newblock


\bibitem[\protect\citeauthoryear{Wong}{Wong}{2021}]%
        {wong_as_nodate}
\bibfield{author}{\bibinfo{person}{Wendy~H. Wong}.}
  \bibinfo{year}{2021}\natexlab{}.
\newblock \showarticletitle{As {U}.{S}. {Capitol} investigators use facial
  recognition, it begs the question: {Who} owns our faces?}
\newblock \bibinfo{journal}{\emph{The Conversation}} (\bibinfo{year}{2021}).
\newblock
\urldef\tempurl%
\url{http://theconversation.com/as-u-s-capitol-investigators-use-facial-recognition-it-begs-the-question-who-owns-our-faces-153253}
\showURL{%
\tempurl}


\bibitem[\protect\citeauthoryear{Wong and Brown}{Wong and Brown}{2013}]%
        {wong_e-bandits_2013}
\bibfield{author}{\bibinfo{person}{Wendy~H. Wong} {and}
  \bibinfo{person}{Peter~A. Brown}.} \bibinfo{year}{2013}\natexlab{}.
\newblock \showarticletitle{E-{Bandits} in {Global} {Activism}: {WikiLeaks},
  {Anonymous}, and the {Politics} of {No} {One}}.
\newblock \bibinfo{journal}{\emph{Perspectives on Politics}}
  \bibinfo{volume}{11}, \bibinfo{number}{4} (\bibinfo{year}{2013}),
  \bibinfo{pages}{1015--1033}.
\newblock


\bibitem[\protect\citeauthoryear{Wu}{Wu}{2010}]%
        {wu_master_2010}
\bibfield{author}{\bibinfo{person}{Tim Wu}.} \bibinfo{year}{2010}\natexlab{}.
\newblock \bibinfo{booktitle}{\emph{The master switch: the rise and fall of
  information empires}}.
\newblock \bibinfo{publisher}{Alfred A. Knopf}, \bibinfo{address}{New York}.
\newblock


\bibitem[\protect\citeauthoryear{Yale}{Yale}{2015}]%
        {Yale2015}
\bibfield{author}{\bibinfo{person}{Elizabeth Yale}.}
  \bibinfo{year}{2015}\natexlab{}.
\newblock \showarticletitle{{The History of Archives: The State of the
  Discipline}}.
\newblock \bibinfo{journal}{\emph{Book History}} \bibinfo{volume}{18},
  \bibinfo{number}{1} (\bibinfo{year}{2015}), \bibinfo{pages}{332--359}.
\newblock


\bibitem[\protect\citeauthoryear{Zuboff}{Zuboff}{2019}]%
        {zuboff_age_2019}
\bibfield{author}{\bibinfo{person}{Shoshana Zuboff}.}
  \bibinfo{year}{2019}\natexlab{}.
\newblock \bibinfo{booktitle}{\emph{The {Age} of {Surveillance} {Capitalism}:
  {The} {Fight} for a {Human} {Future} at the {New} {Frontier} of {Power}}}.
\newblock \bibinfo{publisher}{PublicAffairs}, \bibinfo{address}{New York, NY}.
\newblock


\end{thebibliography}

\end{document}